\documentclass[twocolumn,aps,pra,amsfonts,amssymb]{revtex4}
\usepackage{graphicx}
\usepackage{amsmath}
\usepackage{bm}

\begin{document}

\def\C{{\mathbb{C}}} \def\F{{\mathbb{F}}}
\def\N{{\mathbb{N}}} \def\Q{{\mathbb{Q}}}
\def\R{{\mathbb{R}}} \def\Z{{\mathbb{Z}}}

\def\cH{{\mathcal H}}

\def\bfsigma{\boldsymbol{\sigma}}
\def\bfmu{\boldsymbol{\mu}}

\def\dd{\mathord{\rm d}} \def\Det{\mathop{\rm Det}}
\def\dist{\mathop{\rm dist}} \def\ee{\mathord{\rm e}}
\def\End{\mathord{\rm End}} \def\ev{\mathord{\rm ev}}
\def\id{\mathord{\rm id}} \def\ii{\mathord{\rm i}}
\def\min{\mathord{\rm min}} \def\mod{\mathord{\rm mod}}
\def\prob{\mathord{\rm prob}} \def\tr{\mathop{\rm Tr}}
\def\half{\textstyle\frac{1}{2}} \def\third{\textstyle\frac{1}{3}}
\def\fourth{\textstyle\frac{1}{4}}

\def\vec#1{{\bf{#1}}} \def\vect#1{\vec{#1}}

\def\bra#1{\langle#1|} \def\ket#1{|#1\rangle}
\def\braket#1#2{\langle#1|#2\rangle}
\def\bracket#1#2{\langle#1,#2\rangle} \def\ve#1{\langle#1\rangle}

\def\lR{l^2_{\mathbb{R}}}
\def\RR{\mathbb{R}}
\def\E{\mathbf e}
\def\D{\boldsymbol \delta}
\def\S{{\cal S}}
\def\T{{\cal T}}
\def\dd{\delta}
\def\one{{\bf 1}}
\def\ss{\boldsymbol \sigma}

\title{Towards electron-electron entanglement in Penning traps}
\date{\today}
\author{L. Lamata} \affiliation{Max-Planck-Institut
f\"ur Quantenoptik, Hans-Kopfermann-Strasse 1, D-85748 Garching,
Germany}
\author{D. Porras}\affiliation{Max-Planck-Institut
f\"ur Quantenoptik, Hans-Kopfermann-Strasse 1, D-85748 Garching,
Germany}\author{J.I. Cirac}\affiliation{Max-Planck-Institut f\"ur
Quantenoptik, Hans-Kopfermann-Strasse 1, D-85748 Garching, Germany}
\author{J. Goldman}\affiliation{Department of Physics, Harvard University,
Cambridge, Massachusetts 02138, USA}
\author{G. Gabrielse}\affiliation{Department of Physics, Harvard University,
Cambridge, Massachusetts 02138, USA}

\begin{abstract}
Entanglement of isolated elementary particles other than photons has
not yet been achieved.  We show how building blocks demonstrated
with one trapped electron might be used to make a model system and
method for entangling two electrons.  Applications are then
considered, including two-qubit gates and more precise quantum
metrology protocols.
\end{abstract}

\maketitle

\section{Introduction}

Entanglement is one of the most remarkable features of quantum
mechanics \cite{Nielsen}. Two entangled systems share the holistic
property of nonseparability---their joint state cannot be expressed
as a tensor product of individual states. Entanglement is also at
the center of the rapidly developing field of quantum information
science. A variety of systems have been entangled
\cite{Bouwmeester}, including photons, ions, atoms, and
superconducting qubits. However, no isolated elementary particles
other than photons have been entangled.

It is possible to perform quantum information protocols with electrons in Penning traps, as opposed to ions, even though the former cannot be laser cooled.  This is possible because at low temperature in a large magnetic field (100 mK and 6 T in Harvard experiments \cite{Gabrielse2}) the cyclotron motion radiatively cools to the ground state.  We describe how one could use this mode for quantum information applications since there is sufficiently small coupling to other modes.

In this paper, we describe a possible method for entangling two electrons. The model system and method we investigate are largely based on building blocks already demonstrated with one trapped electron.  On the way to measuring the electron's magnetic moment to $3$ parts in $10^{13}$ \cite{Gabrielse2}, QND methods were used to reveal one-quantum cyclotron and spin transitions between the lowest energy levels of a single electron suspended for months in a Penning trap.  We demonstrate how the two-electron entanglement could make a universal two-qubit gate.  We show how this gate could enable a metrology protocol that surpasses the shot-noise limit, and as an example we consider in detail the requirements for implementing this protocol in a measurement of the electron magnetic moment using two trapped entangled electrons.  The payoffs and requirements for moving from two-electron to $N$-electron entanglement are listed.  Possible applications include quantum simulators \cite{footnote}, analysis of decoherence, and more precise electron magnetic moment measurements using improved quantum metrology protocols.

In Section \ref{Sec2}, we introduce the formalism of two electrons
in a Penning trap. In Section \ref{Sec3}, we obtain a two-electron
gate based on building blocks experimentally demonstrated for one
electron. In Section \ref{Sec4} we identify some applications, including
protocols to entangle the spins of two electrons, a universal
two-qubit gate, spin-cyclotron entanglement generation, and a
quantum metrology protocol with two entangled electrons. In Section
\ref{Sec5} we examine the challenges faced when applying this protocol to existing experimental conditions.  In Section \ref{Sec6} we point out possible extensions to many electrons, and
we present our conclusions in Section \ref{Sec7}.

\section{Two electrons in a Penning trap\label{Sec2}}

A Penning trap \cite{Dehmelt,Gabrielse1} for an electron (charge
$-e$ and mass $m$) consists of a homogeneous magnetic field,
$\vec{B}=B\vec{\hat{z}}$, and an electrostatic quadrupole potential
energy,
\begin{equation}
V_q(\vec{x})= \frac{1}{2}m
\omega_z^2\left(z^2-\frac{\rho^2}{2}\right),
\end{equation}
with $\bm{\rho}=x\vec{\hat{x}}+y\vec{\hat{y}}$. The magnetic field
provides dynamic radial confinement, while the quadrupole potential
gives the axial confinement. The single-particle oscillation
frequencies are the trap-modified cyclotron frequency
$\omega_c^\prime$ (slightly smaller than the cyclotron frequency
$\omega_c = eB/m$), the axial frequency $\omega_z$, the magnetron
frequency $\omega_m$, and the spin precession frequency $\omega_s$.

Quantum control of cyclotron and spin motions was recently achieved
with a single electron  (but not yet with more). Harvard experiments
\cite{Gabrielse2,Gabrielse3} cool an electron's cyclotron motion to
its quantum ground state, using methods that should also work for
more trapped electrons.  Quantum nondemolition (QND) detection and
quantum jump spectroscopy of the lowest cyclotron and spin levels
(with quantum numbers $n$ and $m_s$) \cite{Gabrielse5} cleanly
resolve one quantum transitions to determine $\omega_c^\prime$ and
the anomaly frequency $\omega_a^\prime = \omega_s-\omega_c^\prime$.
These frequencies, with the measured axial frequency $\omega_z$,
determine the dimensionless magnetic moment $g/2$---the magnetic
moment in Bohr magnetons---to an unprecedented level of accuracy.

Crucial to the one-electron measurements, and the multi-particle approach suggested here, are anomaly transitions between the states $|0\rangle|\frac{1}{2}\rangle$ and $|1\rangle|-\frac{1}{2}\rangle$, where the states $|n\rangle|m_s\rangle$ are labeled with their cyclotron and spin quantum numbers.  An applied oscillating electric field drives harmonic axial motion, $z = z_0 \cos{[(\omega_a^\prime + \Delta)}t]$, taking the electron through a ``magnetic bottle'' (MB) gradient,
\begin{equation}
\Delta \vec{B}(\vec{x}) = \beta_2\left[\left(z^2-\frac{\rho^2}{2}\right)\vec{\hat{z}}-z\bm{\rho}\right].
\label{eq:MagneticBottle}
\end{equation}
The electron spin sees the transverse magnetic field oscillating
near $\omega_s$ as needed to flip the spin, at the same time as the
cyclotron motion sees the azimuthal electric field oscillating near
$\omega_c^\prime$, as needed to make a simultaneous cyclotron
transition. The magnetic bottle also enables single-shot
quantum-non-demolition measurements of one-quantum changes in
cyclotron and spin energies, by coupling these to small but
detectable shifts in $\omega_z$ \cite{Gabrielse6}.

We now consider as a model system two electrons in the same Penning trap.  We describe an interaction that will entangle them, we explore possible applications of the entanglement of these two electrons, and then we consider the challenges faced when implementing these techniques in the laboratory.  The Hamiltonian starts with the sum of two single-particle terms,
\begin{eqnarray}
H_i&=&\frac{\hbar \,
\omega_s}{2}\sigma^z_{i}+\frac{[\vec{p}_i+e\vec{A}(\vec{x}_i)]^2}{2m}+V_q(\vec{x}_i).\label{Hamilt1}
\end{eqnarray}
Both the spin and orbital magnetic moments couple to the magnetic bottle, adding
\begin{equation}
H_I =  \sum_{i=1}^2
\mu_B\left[\frac{g}{2}\bm{\sigma}_i+\frac{\vec{L}_i}{\hbar}\right] \cdot \Delta \vec{B}(\vec{x}_i)\label{eq:HI}
\end{equation}
to the Hamiltonian, along with
\begin{eqnarray}
&&V_{12}=\frac{e^2}{4\pi\epsilon_0}\frac{1}{|\vec{x}_1-\vec{x}_2|},
\end{eqnarray}
the Coulomb interaction of the two electrons.

An equilibrium separation of the two electrons can be produced using a so-called ``rotating wall''---a time-dependent oscillating potential in the $x-y$ plane that rotates at angular frequency $\omega \vec{\hat{z}}$ \cite{Huang}.  In the co-rotating reference frame, the effective potential energy is
\begin{eqnarray}
\tilde{V}_q(\vec{x}_i)&=& \frac{1}{2}m {\omega'_\rho}^2 \rho_i^2 +
\frac{1}{2}m \omega_z^2 z_i^2 - \frac{1}{2} m \omega_z^2 \delta
(x_i^2-y_i^2),~~~\label{eq:EffectiveV}
\end{eqnarray}
with
${\omega'_\rho}^2=(\omega_c-\omega)\omega-\frac{1}{2}\omega_z^2$.
The first term on the right in Eq.~(\ref{eq:EffectiveV}) is from the fictitious forces and the transformed quadrupole, the second term is the unchanged axial potential energy, and the third term is the weak rotating wall potential.

Due to the fictitious forces there is radial as well as axial confinement in the rotating frame, provided that $(\omega'_\rho)^2 - \omega_z^2 \delta > 0$.  For small $\delta$ this condition sets a lower limit on the rotating wall frequency, $\omega>\omega_m$.   The total potential energy, $\tilde{V}_q(\vec{x}_1) + \tilde{V}_q(\vec{x}_2) + V_{12}$, is a minimum when the two electrons are diametrically opposed along the direction of weakest confinement.  The corresponding electron equilibrium locations are at $x= \pm x_0/2$, with
\begin{equation}
x_0=\left[\frac{e^2}{2\pi\epsilon_0 m ({\omega'_\rho}^2 - \omega_z^2 \delta)}\right]^{1/3},
\end{equation}
when $\delta>0$ and when the axial confinement is stronger than the
radial confinement, i.e.,\ when $(\omega'_\rho)^2 - \omega_z^2
\delta < \omega_z^2$. For small $\delta$ this condition sets an
upper limit on the rotating wall frequency, $\omega < 3\omega_m$.
The electrons remain near their equilibrium positions because at 100
mK, reachable in principle with current dilution refrigerator
technology \cite{Gabrielse2}, the cyclotron motion cools to its
ground state by synchrotron radiation and the thermal axial
excursion is much less than the radial separation $x_0$.

\section{Two-electron gate\label{Sec3}}

In this model system, the desired entanglement can be achieved by
applying an axial drive at $\omega_a^\prime + \Delta$ in the
presence of a magnetic bottle, as in the one-electron experiments
\cite{Gabrielse2,Gabrielse3}.  The electrons' center of mass
oscillates at this frequency, with
$z_{cm}=z_0\cos[(\omega'_a+\Delta)t]$.  The entanglement arises from
terms $z_{cm} \bm{\rho}_{cm} \cdot (\bm{\sigma}_1 + \bm{\sigma}_2)$
contained in $H_I$ in Eq.~(\ref{eq:HI}), the interaction of the
electrons and the magnetic bottle.  The Hamiltonian can be written
in the well-known Tavis-Cummings form \cite{Tavis} after the
following steps: First, the radial coordinates of the electrons are
written as a sum of center-of-mass (cm) and stretch (st)
coordinates, as in $x_{1,2}=x_{cm}\pm x_{st}/2$, with the cm terms
contributing to producing entanglement.  Second, the cm radial
position is replaced by raising and lowering operators for the cm
magnetron and cyclotron motions, as initially defined for the single
particle case in Ref.~\cite{Gabrielse1}, but with the cm mass $M=2m$
replacing $m$.
\begin{eqnarray}
x_{cm} & = & i \frac{a_{cm,c}-a^\dag_{cm,c}+a_{cm,m}-a^\dag_{cm,m}}{\sqrt{4m/\hbar} ~[\omega_c^2-2\omega_z^2 ]^{1/4}},\\
y_{cm} & = & - \frac{a_{cm,c}+a^\dag_{cm,c}-a_{cm,m}-a^\dag_{cm,m}}{\sqrt{4m/\hbar} ~[\omega_c^2-2\omega_z^2 ]^{1/4}}.
\end{eqnarray}
Third, the rotating wave approximation (RWA) retains only terms that can make anomaly transitions.  Fourth, we switch to the interaction picture with respect to the sum of single particle Hamiltonians for $\delta = 0$.

The resulting Hamiltonian, with the appropriate simplifying choice of the origin of time, is
\begin{eqnarray}
H_I^{\rm gate}=\hbar \Omega
\sum_{i=1,2}(\sigma_i^+a_{cm,c}e^{-i\Delta t}+\sigma_i^-a^{\dag}_{cm,c}e^{i\Delta t}),\label{Tavis}
\end{eqnarray}
which couples the spins and the cm cyclotron motion.

The discarded terms in $H_I$ of Eq.~(\ref{eq:HI}) have been
carefully checked to make sure that they do not produce an
unacceptable decoherence. A thorough analysis including an extension
to $N$ electrons will be included in future work \cite{LamataEtAl}.
The Rabi frequency
\begin{equation}
\Omega=\frac{g}{2}\frac{\mu_B \beta_2z_0 }{\sqrt{4m\hbar} ~[\omega_c^2-2\omega_z^2 ]^{1/4}}.
\end{equation}
Using the parameters from \cite{Gabrielse2}, for an axial oscillation $z_0$ of $100$ $\mu$m, $\Omega/2\pi\sim 10$ Hz.  Unlike the weak drive employed in \cite{Gabrielse2}, this proposal is for a strong drive that induces Rabi flopping.

\section{Applications\label{Sec4}}

We now study applications that would result if this interaction
could be realized.  In an experimental setting, we are subject to
the requirement that a measurement must be performed in a time
shorter than the decoherence time that arises from axial
center-of-mass amplitude fluctuations coupled to $\omega_a^\prime$
via the $z^2$ term in the magnetic bottle \cite{Gabrielse2},
\begin{equation}
t_{\rm dec} \ll  \left[ \omega_a \frac{\beta_2}{B} \frac{k_B T_z}{2m\omega_z^2} \right]^{-1}.
\label{eq:tdec}
\end{equation}
In all the following, our model system includes the assumption that this condition is satisfied.

{\it i)  Entanglement of two electrons in the spin degrees of
freedom.}  This interaction $H_I^{\rm gate}$, along with techniques used in current experiments, makes it possible to obtain a maximally entangled state of the spins of two trapped electrons by performing the following protocol: 1) Begin with state $|\!\!\downarrow\downarrow\rangle|0\rangle_{cm,c}$.  2) Apply a weak resonant cyclotron drive to excite one cyclotron quantum, resulting in the state $|\!\!\downarrow\downarrow\rangle|1\rangle_{cm,c}$.  This state is heralded by projective measurement of the cyclotron center-of-mass mode by detecting shifts in the orthogonal axial oscillation, as customarily done in Harvard experiments \cite{Gabrielse2,Gabrielse3}.  3) Apply an axial drive at $\omega_a^\prime$ for time $t=\pi/(2\sqrt{2}\Omega)$.  This is a $\pi$-pulse with the evolution given by $H_I^{\rm gate}$ (resonant case, $\Delta=0$) of Eq. (\ref{Tavis}).  The population is thereby transferred to the spin-entangled state $(1/\sqrt{2})(|\!\!\uparrow\downarrow\rangle+|\!\!\downarrow\uparrow\rangle)|0\rangle_{cm,c}$.

{\it ii) Two-qubit gates.}  The electron spins are a natural choice for qubits.
When the axial drive is applied far from the anomaly resonance ($\Delta\gg\Omega$), $H_I^{\rm gate}$ takes the form of an effective spin-spin Hamiltonian.  Using adiabatic elimination, this effective off-resonant Hamiltonian for the $|n\rangle_{cm,c}$ subspace is
\begin{equation}
H_I^{\rm off} = \frac{\hbar\Omega^2}{\Delta} \sum_{i,j} \left[ -(n+1)
\sigma^+_{i} \sigma^-_{j}      + n \sigma^-_{i} \sigma^+_{j} \right].\label{offresonantH}
\end{equation}
This Hamiltonian is universal for quantum computation if combined with single-qubit gates.  To perform single-qubit gates in this system it would be necessary to devise a gate that interacts differently with the two electrons, e.g., one with a spatial gradient in the rotating frame.

{\it iii) Entanglement of two electrons in the spin-motional degrees
of freedom.}  By applying $H_I^{\rm gate}$ with a resonant or far
off-resonant drive, it is possible to cause both spin flips and
cyclotron excitations and thereby entangle the spins with the cm
cyclotron motion.  We have verified that one could obtain, among
others, spin-cyclotron GHZ states (see Eq. (\ref{ramsey02e}) below),
useful for quantum metrology protocols.

{\it iv) Precision measurements.}  Standard Ramsey interferometry \cite{Ramsey} allows one to measure a frequency with the maximum precision available with classical means, the shot-noise limit.  A $\pi/2$ pulse, is followed by time $t$ when no drive is applied, and then by a second $\pi/2$ pulse.

A measurement precision improved by a factor $\sqrt{N}$ is possible when a maximally entangled state of $N$ particles is used with quantum metrology (QM) protocols \cite{Bollinger,Lloyd}.  For example, the two particle state
\begin{equation}
|\Psi\rangle= \frac{
|\!\!\uparrow\uparrow\rangle|0\rangle_{cm,c}+|\!\!\downarrow\downarrow\rangle|2\rangle_{cm,c}
}{\sqrt{2}}, \label{ramsey02e}
\end{equation}
can yield a measurement uncertainty $\delta
\omega'_a=1/(2\sqrt{Tt}),$ where $t$ is the time for a single
measurement and $T$ the total time of the experiment.  This
so-called Heisenberg limit is a $\sqrt{2}$ improvement over the
shot-noise limit for two electrons, and is the lowest uncertainty
allowed by the laws of quantum mechanics. Here we show how in
principle the Heisenberg limit could be reached by making use of
$H_I^{\rm gate}$ in Eq.~(\ref{Tavis}).  In fact, the gain in
precision is a factor of 2 compared to a single-particle Ramsey
experiment: one factor of $\sqrt{2}$ from classical counting
statistics since we are using two particles, and an additional
factor of $\sqrt{2}$ from using an entangled state with the given
protocol.  Achieving this gain would be of modest significance in
its own right. It would also be an important first step toward
extending the protocol to more particles, with the uncertainty
decreasing by as much as $1/N$, the fundamental Heisenberg limit, a
significant improvement over the shot-noise limit of $1/\sqrt{N}$.

\section{Two-electron metrology protocol\label{Sec5}}

We now consider the two-electron metrology protocol in more detail to examine the essential elements and feasibility of performing this protocol in typical laboratory conditions.  The effective ``$\pi/2$ pulse'' needed \cite{Bollinger} to produce the maximally entangled state in Eq.~(\ref{ramsey02e}) (up to an overall phase) is a sequence of three pulses.  These are applied to an initial state
\begin{equation}
|\Psi(0)\rangle=|\!\!\uparrow\uparrow\rangle|0\rangle_{cm,c}
\end{equation}
that is prepared much as the one-particle state $|\!\!\uparrow\rangle|0\rangle_c$ is currently prepared.  First is a
resonant pulse applied for a duration of $t_0 = 1.027/\Omega$.  Second is a far off-resonant pulse applied for a duration of $t_1 = \pi\Delta/(6\Omega^2)$.  Third is a resonant pulse of duration $t_2 = 1.140/\Omega$.

In this variation of the Ramsey method the system next evolves freely for time $t$.  The second effective ``$\pi/2$'' pulse that is next applied is also a three pulse sequence.  First is a resonant pulse applied for time $1.425/\Omega$.  Second is an off-resonant pulse with a duration of $\pi\Delta/(6\Omega^2)$.  Third is a resonant pulse with a duration of $1.538/\Omega$.  The number of center-of-mass cyclotron excitations, $a^\dag_{cm,c}a_{cm,c}$, is then measured by detecting shifts in the center-of-mass axial frequency as is currently done for one particle \cite{Gabrielse5}.

\begin{figure}[t!]
\begin{center}
\includegraphics[width=\linewidth]{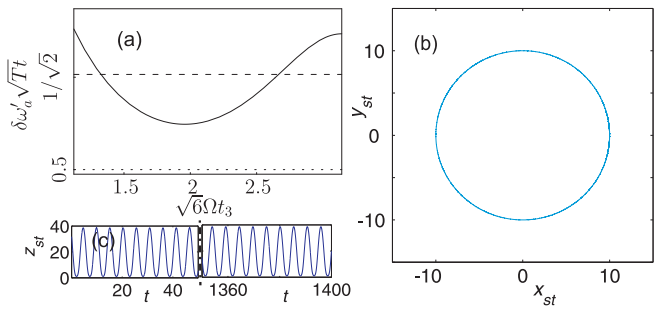}
\end{center}
\caption{(color online). (a) $\delta\omega'_a\sqrt{Tt}$ as a function of $\sqrt{6}\Omega t_3$, for a partially entangled state obtained by a single application of $H_I^{\rm gate}$, Eq. (\ref{Tavis}) (resonant case)  during a time $t_3$, that for $t_3\simeq 2/(\sqrt{6}\Omega)$ obtains an improvement over the best classical strategy (shot-noise limit, dashed). Shown is also Heisenberg limit, dotted. (b,c) Classical trajectory of the relative coordinates $x_{st}$, $y_{st}$, $z_{st}$.  All the spatial coordinates are relative to $\sqrt{\hbar/m\omega_z}$, and time to $1/\omega_z$.} \label{Fig1}
\end{figure}

A substantial fraction of the uncertainty reduction below the shot-noise limit can be attained even if the off-resonant part of the interaction sequence is omitted.  A single resonant pulse of duration $t_3=0.76/\Omega$ is applied, followed by a time $t$ during which no drive is applied, and then a resonant pulse of duration $2\pi/(\sqrt{6} \Omega )-t_3=1.80/\Omega$.   Fig. \ref{Fig1}(a) shows how the realized uncertainty (solid curve) is lower than the shot-noise limit (dashed curve) but higher than the Heisenberg limit (dotted curve) for the optimal choice of $t_3$.

The Ramsey method has not been used for the one-particle measurements.  Instead, a much weaker drive applied over a much longer time was used to drive anomaly transitions at $\omega_a^\prime$.  When performing an experiment using the above protocol on a similar experimental system, the drive strength, and hence the axial excursion of the electrons, must be substantially increased, so it will be necessary to ensure that the resonant frequency does not shift due to the increased oscillation amplitude and drive strength.

The required drive amplitude is determined by the requirement that a
single measurement be performed in much less than $t_{\rm dec}$ of
Eq.~\ref{eq:tdec}.  At 300 mK, under the conditions realized thus
far with one electron, this decoherence time is 2.3 seconds.  We
thus assume that we need the duration of each measurement to be 10\%
of this limit (about 0.23 s).  The effective $\pi/2$ pulses must be
completed in a much shorter time; we will use $t/10 = 23$ ms for the
following estimates.  Completing the second pulse sequence in 23 ms
requires a Rabi frequency $\Omega/(2\pi) = 57$ Hz.  This corresponds
to driving the electron to an axial oscillation amplitude of nearly
0.5 mm for the one-electron parameters that have been realized,
requiring a 24 V peak-to-peak driving voltage superimposed upon the
100 V trapping potential.  Such a large drive and oscillation
amplitude is not excluded in principle since very large oscillation
amplitudes have been employed for the detection of a single
electron.  However, this amplitude is about $10^3$ times  larger
than the amplitude used to carry out quantum jump spectroscopy,
during which time it is very important to avoid frequency shifts
caused by magnetic and electrical gradients in the trap.  Careful
investigations of the effects of the large oscillatory potential,
and the large axial amplitude, will be required to make sure that
$\omega_a^\prime$ is not unacceptably modified.

Similarly, the duration of a measurement must in practice be less than the cyclotron damping time.  Cavity-inhibited spontaneous emission has produced damping times as long as 15 seconds \cite{GabrielseDehmelt}, but damping times less than a second have been used to investigate systematic effects, so care must be taken here.  Coherent modifications of the damping time must also be investigated.

We have analyzed all of the neglected terms from the interaction of
the drive with the two electron system, including center-of-mass and
stretch normal modes. The frequency splitting of center-of-mass and
stretch cyclotron mode, of tens of kilohertz, is large enough such
that the coupling with the stretch mode will negligibly affect the
gate. These terms can indeed be neglected without effect if the
couplings $\tilde{\Omega}$ are weak and detunings $\tilde{\Delta}$
are large. Each such term gives a probability of an unintended
change in the quantum state proportional to
$(\tilde{\Omega}/\tilde{\Delta})^2$. All of the terms neglected from
$H_I$ have $(\tilde{\Omega}/\tilde{\Delta})^2 < 10^{-4}$, and in
most cases many orders of magnitude smaller.  Thus, the protocol
will be negligibly affected by the neglected terms.  In addition,
numerical simulations that include the full Coulomb interaction (not
just the quadratic expansion) confirm that the internal stretch
motion undergoes stable oscillations even for very large axial
amplitudes [see Fig. \ref{Fig1}(b,c)].

\section{Requirements for extending to $N$ electrons\label{Sec6}}
For both metrology and quantum information processing, much bigger
gains could be achieved by entangling more than two electrons.  Up
to many millions of electrons can be simultaneously stored in a
Penning trap, though coherent control has only been obtained so far
with one trapped electron and for the center-of-mass motion of many
particles.  Extending these methods to many electrons presents
several challenges.  First, more than two electrons must be cooled
to form a Wigner crystal, and the rotation frequency selected to
make a planar crystal.  Second, new pulse sequences must be devised
to make effective $\pi/2$ pulses to entangle the electrons. Although
the shot noise limit could likely be surpassed, whether resonant and
off-resonant anomaly drives could maximally entangle $N$ electrons
remains to be proven.  Third, methods to compensate for the finite
extent of the electron crystal must be devised, such as the
inhomogeneous broadening from spin frequencies that differ due to
the $\rho^2$ term in the magnetic bottle
[Eq.~(\ref{eq:MagneticBottle})], unless spins on the same circular
orbit could be utilized.  Fourth, the crystal's normal mode spectrum
must be derived, and modes with frequencies near $\omega_{cm,c}$
must be checked to ensure that their coupling does not produce
unacceptable decoherence errors.  These questions will be discussed
in detail in future work \cite{LamataEtAl}.

\section{Conclusions\label{Sec7}}
In conclusion, we have shown how two electrons could be entangled
using a model system whose building blocks have already been
demonstrated with one electron.  An experimental investigation is
now needed to determine the completeness of the model system.  An
experimental realization of two-electron entanglement would be the
significant first entanglement of isolated elementary particles
other than photons.  A universal two-qubit gate for quantum
computing could then be demonstrated, and it may be feasible to
improve the measurement precision for the electron magnetic moment
by a factor of two.  In addition, the realization of two-electron
entanglement would be a first step towards learning about and
realizing entangled states of larger numbers of electrons for
quantum simulators, and much larger precision gains in magnetic
moment measurements using quantum metrology protocols.

\section{Acknowledgements}
L.L.\ and G.G.\ acknowledge funding by Alexander von Humboldt Foundation, and L.L.\ the  support by Spanish MICINN project FIS2008-05705/FIS. We acknowledge support by European IP SCALA and DFG Excellenzcluster NIM, and by the US NSF.


\begin{thebibliography}{99}
\bibitem{footnote} An analysis of a quantum processor and quantum
simulator based on arrays of microtraps containing one electron each
was performed \cite{Tombesi}.
\end{thebibliography}
\end{document}